\documentclass[12pt,a4paper]{article}

\usepackage{epsfig}

\newcommand{\eqpt}{\hspace{6pt}.}           % punctuation in formula
\newcommand{\eqcm}{\hspace{6pt},}

\newcommand{\eqref}[1]{(\ref{#1})}          % equation numbers

\newcommand{\arreq}{\hspace{0.4em} = \hspace{0.4em}}
\newcommand{\GeV}{\mbox{\ GeV}}             % units
\newcommand{\fbarn}{\mbox{\ fb}}
\newcommand{\pbarn}{\mbox{\ pb}}

\newcommand{\obs}{{\cal O}}                 % observables
\newcommand{\mean}{\bar{\obs}}              % their mean values
\newcommand{\sig}[1]{\hat{\sigma}_{#1}}     % normalised coefficients

\newcommand{\CP}{{\it C P}}                 % discrete symmetries
\newcommand{\CPT}{{\it C P \! \tilde{T}}}

\newcommand{\bvec}[1]{\mathbf{#1}}          % boldface for vectors

\newcommand{\wein}{\sin^2 \theta_W}         % Weinberg angle

\newcommand{\re}{\textrm{Re}}               % Real part

\begin{document}

% title page
\setcounter{page}{0}
\pagestyle{empty}
\begin{flushright}
{HD--THEP--97--03 \\ CPTH--S494--0197}
\end{flushright}
\vspace{\baselineskip}
\begin{center}
{\Large ANOMALOUS THREE GAUGE BOSON \\}
\smallskip
{\Large  COUPLINGS IN $e^+e^-\to W^+W^-$ AND \\}
\smallskip
{\Large  ``OPTIMAL'' STRATEGIES FOR \\}
\smallskip
{\Large THEIR MEASUREMENT \\}
\vspace{2\baselineskip}
M. Diehl\footnote{email: diehl@orphee.polytechnique.fr} \\[\baselineskip]
\textit{Department of Applied Mathematics and Theoretical Physics \\
  Silver Street, Cambridge CB3 9EW, Great Britain \\[\baselineskip]
  present address: \\ Centre de Physique Th\'eorique\footnote{Unit\'e
    propre 14 du CNRS} \\ Ecole Polytechnique, F-91128 Palaiseau
  Cedex, France} \\[\baselineskip]
and \\[\baselineskip]
O. Nachtmann\footnote{email: O.Nachtmann@thphys.uni-heidelberg.de}
\\[\baselineskip]
\textit{Institut f\"ur Theoretische Physik \\ Philosophenweg 16,
  D-69120 Heidelberg, Germany}
\vspace{2\baselineskip}
\begin{abstract}
  We show how one can measure anomalous $WWZ$- and
  $WW\gamma$-couplings with minimal statistical error using integrated
  observables, without having to assume that the anomalous couplings
  are small. We propose a parametrisation of these couplings which is
  well suited for the extraction of both single and many parameters,
  and which leads to a very simple form of the integrated cross
  section, from which additional information on the couplings can be
  obtained.
\end{abstract}
\end{center}

\newpage

\pagestyle{plain}

\section{Introduction}
\label{sec:intro}

The direct and precise measurement of the self-coupling between the
electroweak gauge bosons in $W$-pair production will be a crucial step
in testing the standard model of electroweak interactions and
searching for physics beyond it. It will form an important part of the
physics programme at LEP2 and at a planned linear $e^+ e^-$-collider
(LC). As is well known there are three diagrams at tree level that
contribute to the amplitude of $e^+ e^- \to W^+ W^-$ in the standard
model, one with $t$-channel neutrino exchange and the other two with a
$\gamma$ or $Z$ in the $s$-channel, involving the vertices $WW\gamma$
and $WWZ$. One can parametrise the corresponding vertex functions in
order to quantify the couplings and to compare them with their form in
the standard model. In the most general form respecting Lorentz
covariance each vertex involves seven complex form factors
\cite{Hagiwa}, three of which give couplings that violate $\CP$
symmetry.

Without further physical assumptions one is thus left with 28 real
parameters whose simultaneous extraction in one experiment looks quite
hopeless. Given the limited event statistics expected at both LEP2 and
the LC one will only obtain meaningful errors on a reduced number of
coupling parameters at one time. This may be achieved by imposing
certain constraints on the full set of coupling constants; various
suggestions for such constraints based on symmetry considerations have
been made in the literature \cite{LEP2:workshop,Bilenky}. One must
however keep in mind that experimental values or bounds on couplings
that have been obtained with particular constraints cannot be
converted into results without constraints or with different ones; the
information lost by assuming relations between couplings cannot be
retrieved. Although imposing such constraints is certainly legitimate
and can be useful we stress that a data analysis with independent
couplings will be valuable, both from the point of view of model
independence and the capability to compare results of different
experiments.

We remark that of course one can also give (reasonably small) errors
on \emph{single or few} couplings in a multi-parameter analysis. In
this paper we propose a parametrisation of the couplings which is well
adapted to this end, the statistical errors on the different measured
parameters being approximately uncorrelated.

We will work in the framework of optimal observables, a way to extract
unknown coupling parameters introduced for the case of one parameter
in \cite{Atwood,Davier} that has since been used for various reactions
\cite{OtherOpt,Papadop,LEP1:tau}.  General aspects of this method, in
particular its extension to an arbitrary number of parameters, as well
as its application to $W^+ W^-$ production were discussed in
\cite{OldObs}.  In this paper we investigate again the reaction
$e^+e^-\to W^+W^-$. We concentrate here on the decay channels, where
one $W$ decays hadronically and the other into an electron or muon and
its neutrino.  Calculated with the Born level cross section of the
standard model the statistics of these channels is about 3000 events
for a collision energy of $\sqrt{s} = 190 \GeV$ and $500 \pbarn^{-1}$
integrated luminosity, which are typical planned LEP2 parameters, and
about 22000 events with $10 \fbarn^{-1}$ at $\sqrt{s} = 500 \GeV$,
which might be achieved at the LC.

A complementary source of information is the integrated cross section,
which is a quadratic function of the triple gauge couplings. The
combination of information from the total event rate and from
observables that make use of the detailed distribution in the final
state has for example been used in \cite{LEP1:bbar}, where $\CP$
violation in the decay $Z \to b \bar{b} g$ was investigated.

In sec.~\ref{sec:method} of this paper we will further develop some
aspects of the method of optimal observables, in particular we will
show how to apply it without the linear approximation in the coupling
parameters that was used in \cite{OldObs}. In sec.~\ref{sec:diagon} we
then propose a parametrisation of the couplings that simultaneously
diagonalises certain matrices connected with our observables and with
the integrated cross section. These parameters achieve two goals:
their quadratic contribution to the total cross section is a simple
sum of squares and the covariance matrix of the corresponding optimal
observables is diagonal. The methods which we use for this purpose are
borrowed from the theory of small oscillations of a system with $f$
degrees of freedom (cf.\ e.g.~\cite{Gold}). Our parameters correspond
to ``normal coordinates'' and their use in an experimental analysis
should in our view present several advantages. We give some numerical
examples for $W$-pair production at LEP2 and the LC in
sec.~\ref{sec:numeric} and make some further remarks on how our
proposal might be implemented in practice in sec.~\ref{sec:practice}.
The last section of this paper gives a summary of our main points.

\section{Optimal observables: analysis beyond leading order}
\label{sec:method}

The method of optimal observables has previously been presented in the
approximation that the couplings to be extracted are sufficiently
small to allow for a leading order Taylor expansion of various
expressions. Here we show how to use it beyond this approximation.

Let us denote by $g_i$ the real and imaginary parts of the $WW\gamma$
and $WWZ$ form factors minus their values in the standard model at
tree level. As the amplitude of our process is linear in these
couplings we can write the differential cross section as
\begin{equation}
  \label{diffXsection}
  \frac{d \sigma}{d\phi} = S_0(\phi) + \sum_{i} S_{1,i}(\phi) \, g_i +
  \sum_{ij} S_{2,ij}(\phi) \, g_i g_j  \eqcm
\end{equation}
where $S_{2,ij}(\phi)$ is a positive semidefinite symmetric matrix.
$\phi$ collectively denotes the set of measured phase space variables.
The integrated cross section is
\begin{equation}
  \label{intXsection}
  \sigma = \sigma_0 \left( 1 + \sum_{i} \sig{1,i} \, g_i +
    \sum_{ij}\sig{2,ij} \, g_i g_j \right)
\end{equation}
with the standard model cross section $\sigma_0 = \int d\phi \,
S_0(\phi)$ and coefficients
\begin{equation}
  \label{XsectionCoeffs}
  \sig{1,i} = \frac{1}{\sigma_0}{\int d\phi \, S_{1,i}}(\phi) \eqcm
  \hspace{2em}
  \sig{2,ij} = \frac{1}{\sigma_0}{\int d\phi \, S_{2,ij}}(\phi) \eqpt
\end{equation}
The idea of using integrated observables is to define suitable
functions $\obs_i(\phi)$ of the phase space variables and to extract
the unknown couplings from their measured mean values $\mean_i$. Let
us give the details. From \eqref{diffXsection} and \eqref{intXsection}
we obtain the expectation value $E[\obs_i]$ of $\obs_i$ as
\begin{equation}
  \label{expect}
  E[\obs_i] - E_0[\obs_i] = \frac{\displaystyle \sum_{j} c_{ij} \, g_j
    + \sum_{jk} q_{ijk} \, g_j g_k}{\displaystyle 1 + \sum_{j}
    \sig{1,j} \, g_j + \sum_{jk} \sig{2,jk} \, g_j g_k}
\end{equation}
with the standard model expectation value $E_0[\obs_i] = (\int d\phi
\, \obs_i S_0) / \sigma_0$ and coefficients
\begin{eqnarray}
  \label{obsCoeffs}
  c_{ij} &=& \frac{1}{\sigma_0} \int d\phi \, \obs_i S_{1,j} -
  E_0[\obs_i] \, \sig{1,j}  \eqcm \nonumber \\
  q_{ijk} &=& \frac{1}{\sigma_0} \int d\phi \, \obs_i S_{2,jk} -
  E_0[\obs_i] \, \sig{2,jk} \eqpt
\end{eqnarray}
We remark in passing that the coefficients in \eqref{expect} can be
written in a compact form as
\begin{eqnarray}
  \label{obsCoeffCompact}
  c_{ij} &=& V_0[\obs_i \, , \; S_{1,j} /S_0] \eqcm \hspace{3em}
  q_{ijk} \arreq V_0[\obs_i \, , \; S_{2,jk} /S_0] \eqcm \nonumber \\
  \sig{1,j} &=& E_0[S_{1,j} /S_0] \eqcm \hspace{4.4em}
  \sig{2,jk} \arreq E_0[S_{2,jk} /S_0] \eqcm
\end{eqnarray}
where $V_0[f, g] = E_0[f g] - E_0[f] \, E_0[g]$ is the covariance of
$f(\phi)$ and $g(\phi)$ in the standard model. Note that $\sig{2,jk}$
is symmetric and positive definite, whereas $q_{ijk}$ as a matrix in
$j$ and $k$ is symmetric but in general indefinite.

An estimation of the couplings can now be obtained by solving the
system \eqref{expect} with $E[\obs_i]$ replaced by the mean values
$\mean_i$,
\begin{equation}
  \label{mean}
  \mean_i - E_0[\obs_i] = \frac{\displaystyle \sum_{j} c_{ij} \, g_j
    + \sum_{jk} q_{ijk} \, g_j g_k}{\displaystyle 1 + \sum_{j}
    \sig{1,j} \, g_j + \sum_{jk} \sig{2,jk} \, g_j g_k}  \eqcm
\end{equation}
provided of course one has $n$ observables for $n$ unknown couplings.
When the system \eqref{mean} is linearised in the $g_i$ it is easily
solved by inversion of the matrix $c_{ij}$. One is however not
constrained to do so and can instead solve the exact set of equations
\eqref{mean}. By multiplication with the denominator it can be
rearranged to a coupled set of quadratic equations in the $g_i$ and
will in general have several solutions. Some of these may be complex
and thus ruled out, but from the information of the $\mean_{i}$ alone
one cannot tell which of the remaining real ones is the physical
solution.  We will come back to this point.

The measured mean values $\mean_{i}$ are of course only equal to the
$E[\obs_i]$ up to systematic and statistical errors. We only consider
the latter here, which are given by the covariance matrix
$V(\obs)_{ij}$ of the observables $\obs_i$ divided by the number $N$
of events in the analysis. To convert the errors on the observables
into errors on the extracted couplings we use the quantity
\begin{equation}
  \label{chi}
  \chi^2(g) = \sum_{ij} \left( \mean_i - E[\obs_i] \right) N
  V(\obs)^{-1} {}_{ij} \left( \mean_j - E[\obs_j] \right) \eqcm
\end{equation}
which depends on the $g_i$ through the $E[\obs_i]$ given in
\eqref{expect}. Solving \eqref{mean} is tantamount to minimising
$\chi^2$ with $\chi^2_{\it min} = 0$, and a confidence region on the
couplings is as usual given by
\begin{equation}
  \label{confidence}
  \chi^2(g) - \chi^2_{\it min} \le \textit{const.}
\end{equation}
with the constant determined by the desired confidence level.

There are several possible choices for the covariance matrix $V(\obs)$
in \eqref{chi}. It can be
\begin{enumerate}
\item determined from the measured distribution of the observables
  $\obs_i$,
\item calculated from the differential cross section
  \eqref{diffXsection}, taking for the $g_i$ the values extracted in
  the measurement,
\item calculated for vanishing couplings $g_i$,
\item calculated as a function of the couplings.
\end{enumerate}
Choices 1.\ and 2.\ should lead to the same results in the limit of
large $N$ where the statistical errors on the measured $V(\obs)_{ij}$
and $g_i$ become small. Comparison of the covariance matrices obtained
by these two methods might indeed be helpful to rule out unphysical
solutions of \eqref{mean}. Choice 3.\ in turn will be a good
approximation of 2.\ if the couplings are small enough. We consider
possibility 4.\ as the least practical one, except maybe for the case
of one coupling. For several couplings the expression of
$V(\obs)_{ij}$ as a function of the $g_i$ involves tensors of rank up
to four and is even more complicated than the one for the expectation
values \eqref{expect}, and the inverse matrix is yet more clumsy. For
this reason we will discard choice 4.\ in the following.

In \cite{OldObs} we considered an analysis at leading order in the
$g_i$, where one uses the linearised form of \eqref{mean} to estimate
the couplings:
\begin{equation} \label{linearEstim}
  \hat g_j = \sum_k c^{-1} {}_{jk} \left( \mean_k - E_0[\obs_k]
\right)  \eqpt 
\end{equation}
Correspondingly the linear approximation of \eqref{expect} is used in
the expression \eqref{chi} of $\chi^2$ which then reads
\begin{equation} \label{linearChi}
  \chi^2(g) = \sum_{ij}
  (\hat g_i - g_i) \, V(g)^{-1} {}_{ij} \, (\hat g_j - g_j)  \eqcm
\end{equation}
where
\begin{equation}
  \label{couplingsCov}
  V(g)^{-1} = N \, c^T \cdot V(\obs)^{-1} \cdot c
\end{equation}
is the inverse covariance matrix of the estimated couplings
\cite{OldObs}. As one works to leading order in the $g_i$ one can
approximate $V(\obs)$ by its value for zero couplings, i.e.\ choose
possibility 3.\ above.

The confidence regions $\chi^2(g) \leq \textit{const.}\,$ for the
measured couplings are then ellipsoids in the space of the $g_i$ with
centre at $(\hat g_i)$. The optimal observables
\begin{equation}
  \label{optimal}
  \obs_i(\phi) = \frac{S_{1,i}(\phi)}{S_0(\phi)}
\end{equation}
discussed in \cite{OldObs} have the property that to leading order the
statistical errors on the estimated couplings are the smallest
possible ones that can be obtained with \emph{any} method, including
e.g.\ a maximum likelihood fit to the full distribution of $\phi$
given by the differential cross section \eqref{diffXsection}.

Note that one can still use the linearised expressions
\eqref{linearEstim} and \eqref{linearChi} in an analysis beyond
leading order. The error $\chi^2(g)\leq1$ on the couplings will be
given by an ellipsoid with defining matrix \eqref{couplingsCov}, where
$V(\obs)$ is the covariance matrix at the actual values of the
couplings. These errors will in general no longer be optimal, so that
when the leading order approximation is not good one might obtain
better errors with a different choice of observables.  More
importantly, however, the extracted values of the couplings are
biased: averaged over a large number of experiments the measured
couplings differ from the actual ones by terms quadratic in the $g_i$.

If instead one uses the full expressions \eqref{expect}, \eqref{mean}
and \eqref{chi} one has no bias on the extracted coupling parameters,
provided the number $N$ of events in the analysis is large enough. Let
us see if we can find optimal observables for this case. To this end
we expand the differential cross section around some values
$\tilde{g}_i$ of the couplings:
\begin{equation}
  \label{newDiffXsection}
  \frac{d \sigma}{d\phi} = \widetilde{S}_0 + \sum_{i}
  \widetilde{S}_{1,i}(\phi) \cdot ( g_i - \tilde{g}_i ) + \sum_{ij}
  \widetilde{S}_{2,ij}(\phi) \cdot ( g_i - \tilde{g}_i ) ( g_j -
  \tilde{g}_j )  \eqpt
\end{equation}
The corresponding zeroth order cross sections and mean values are
$\widetilde{\sigma}_0 = \int d\phi \, \widetilde{S}_0$ and
$\widetilde{E}_0[\obs_i] = (\int d\phi \, \obs_i \widetilde{S}_0) /
\widetilde{\sigma}_0$, respectively. We then can re-express
$E[\obs_i]$ in \eqref{expect}, replacing $g_i$ with $g_i -
\tilde{g}_i$, $E_0$ with $\widetilde{E}_0$, and using new coefficients
$\tilde{c}_{ij}$ etc.\ constructed as in \eqref{XsectionCoeffs},
\eqref{obsCoeffs}. Making the same replacements in \eqref{mean} we
have an alternative set of equations to extract the coupling
parameters.

It can be shown that for sufficiently large $N$ the confidence regions
obtained from \eqref{chi}, \eqref{confidence} in a nonlinear analysis
are again ellipsoids given by $\chi^2(g) \leq \textit{const.} \,$ One
can then write $\chi^2(g)$ as in \eqref{linearChi}, but with
$V(g)^{-1}$ of \eqref{couplingsCov} replaced by
\begin{equation}
  \label{newCouplingsCov}
  V(g)^{-1} = N \, \tilde{c}^{\; T} \cdot V(\obs)^{-1}
  \cdot \tilde{c}  \eqcm
\end{equation}
where $\tilde{c}$ corresponds to an expansion \eqref{newDiffXsection}
of $d \sigma / d\phi$ about the \emph{actual} values of the couplings.
The main point of the argument is that for large $N$ the statistical
errors on the $\mean_i$ become small, so that the extracted couplings
will be sufficiently close to the actual ones to allow for a
linearisation of \eqref{expect} and \eqref{mean}, cf.~\cite{Recipes},
p.695, and \cite{Avni}.

Finally one can construct new observables
\begin{equation}
  \label{newOptimal}
  \widetilde{\obs}_i(\phi) =
  \frac{\widetilde{S}_{1,i}(\phi)}{\widetilde{S}_0(\phi)}
\end{equation}
from \eqref{newDiffXsection}. They will be optimal, i.e.\ have minimum
statistical error if the $\tilde{g}_i$ are equal to the actual values
of the $g_i$. In the appendix we show that, up to linear
reparametrisations given in \eqref{reparam}, this is the only set of
$n$ integrated observables that measures the $n$ couplings with
minimum error. There is hence no choice of observables that would be
optimal for \emph{all} values of the actual coupling parameters. As
these are unknown one can in practice not write down the truly
``optimal'' observables, but our argument tells us how one can improve
on the choice in \eqref{optimal} if one has some previous estimates
$\tilde{g}_i$ of the couplings (cf.\ also \cite{Davier}). One may then
choose to perform a leading order analysis as described above,
linearising about $g_i = \tilde{g}_i$ instead of $g_i = 0$. A
practical way to proceed could be to estimate the parameters $g_i$ at
first using the linearised method around $g_i=0$. Suppose this gives
as best estimate some values $\hat g_i$. Then in a second step one
could set $\tilde g_i=\hat g_i$ and use the linearised method around
$\tilde g_i$ to improve the estimate etc.

At this point we wish to comment on the ``optimal technique'' for
determining unknown parameters in the differential cross section that
has been proposed in \cite{Grzad}. The ``weighting functions''
$w_i(\phi)$ there depend on the actual values of the parameters one
wants to extract and are thus not ``observables''. Only if one sets
the unknown parameters in the $w_i(\phi)$ equal to some previous
estimates of them can one use these functions to weight individual
events; the better these estimates are the more sensitive the
functions will be.  If one does this then the set $w_i(\phi)$ is
equivalent to our observables \eqref{newOptimal} defined for some
estimates $\tilde{g}_i$ of the coupling parameters.

We finally remark that if $N$ is not large enough the statistical
errors on the mean values $\mean_i$ and thus on the measured couplings
might be so large that they lead into a region where a linearisation
of \eqref{expect} is not a good approximation. The covariance matrix
$V(g)$ is then no longer given by \eqref{newCouplingsCov}. Moreover
the errors on the couplings might be asymmetric and the shape of the
confidence region defined by \eqref{chi}, \eqref{confidence} very
different from an ellipsoid, so that knowledge of $V(g)$ is not
sufficient to estimate the errors on the $g_i$. In such a case we
cannot say on general grounds how sensitive our observables are.
Incidentally this also holds for other extraction methods such as
maximum likelihood fits, whose optimal properties are realised in the
limit $N\to\infty$. If one is rather far from this limit the
sensitivity of a method will have to be determined by other means,
e.g.\ by detailed Monte Carlo simulations.

The method we have outlined can of course also be applied if one
chooses to reduce the number of unknown parameters  by
imposing certain linear constraints on the couplings. One may still
use the observables \eqref{optimal} corresponding to the \emph{full}
set of couplings but minimise $\chi^2$ in \eqref{chi} for the
\emph{reduced} set; in this case one can of course not take choice 2.\
for $V(\obs)$. In general $\chi^2_{\it min}$ is then different from
zero and its value indicates to which extent the particular
constraints on the couplings are compatible with the data. If $N$ is
large enough $\chi^2_{\it min}$ follows in fact a
$\chi^2$-distribution with $n - m$ degrees of freedom for $n$
observables and $m$ independent couplings so that its value can be
converted into a confidence level \cite{Recipes}.

We conclude with a remark on the use of optimal observables in
practice. A realistic data analysis will not be good enough if the
Born approximation of the differential cross section
\eqref{diffXsection} is used. Both higher-order theoretical
corrections, such as initial state radiation and the finite $W$ width,
and experimental effects like detection efficiency and resolution will
modify the observed distribution of the phase space parameters $\phi$.
If they are taken into account in the determination of the
coefficients in \eqref{expect}, \eqref{mean} and of the covariance
matrix $V(\obs)$ they will \emph{not} lead to any bias in the
extraction of the couplings and their errors. While this will
presumably be done with sets of generated events and might be
computationally intensive one still has to determine only a rather
limited number of ``sensitivity'' constants.  On the other hand one
needs to know the observables $\obs_i(\phi)$ of \eqref{optimal} as
functions over the entire experimental phase space, so that the
expressions of $S_0$ and $S_{1,i}$ used to construct them will in
practice be taken from a less sophisticated approximation to the
actual distribution of $\phi$ in order to keep them manageable.  The
observables are then no longer optimal, and it will depend on the
individual case which approximations of $S_0$, $S_{1,i}$ are good
enough to obtain observables with a sensitivity close to the optimal
one.

\subsection{Discrete symmetries}
\label{sec:symmetries}

In \cite{OldObs} it was shown how with a suitable combination of all
semileptonic $W^- W^+$ decay channels one can define observables that
are either even or odd under the discrete transformations $\CP$ and
$\CPT$, where $C$ denotes charge conjugation, $P$ the parity
transformation, and $\tilde{T}$ the ``naive'' time reversal operation
which flips particle momenta and spins but does not interchange
initial and final state. Under the conditions on the experimental
setup and event selection spelt out in \cite{OldObs} we have two
important symmetry properties:
\begin{enumerate}
\item A $\CP$ odd observable can only have a nonzero expectation value
  if $\CP$ symmetry is violated in the reaction.
\item If the expectation value of a $\CPT$ odd observable is nonzero
  the transition amplitude must have an absorptive part whose phase
  must satisfy certain requirements in order to give an interference
  with the nonabsorptive part of the amplitude.
\end{enumerate}
We assume in this analysis that any nonstandard physics in the
reaction is due to the triple gauge vertices. In the standard model
one needs at least two loops to violate $\CP$; to a good accuracy the
triple gauge couplings are therefore the only possible source of $\CP$
violation. For our process, i.e.\ $e^+ e^-$ annihilation into four
fermions, an absorptive part that satisfies the requirements mentioned
in point 2.\ will appear in the Standard Model already at
next-to-leading order in the electroweak fine structure constant,
either through nonresonant diagrams or through loop corrections. To
leading order, however, they are only due to the imaginary parts of
triple gauge couplings.

In this approximation the optimal observables \eqref{optimal} are
$\CP$ even (odd) if they correspond to $\CP$ conserving (violating)
couplings, and $\CPT$ even (odd) if they correspond to the real
(imaginary) parts of form factors. The coefficient matrix $c_{i j}$ is
then block diagonal in four symmetry classes of observables and
three-boson-couplings:
\begin{enumerate}
\item[$(a):$] $\CP$ and $\CPT$ even
\item[$(b):$] $\CP$ even and $\CPT$ odd
\item[$(c):$] $\CP$ odd and $\CPT$ even
\item[$(d):$] $\CP$ and $\CPT$ odd.
\end{enumerate}

In the leading order analysis one thus can treat these four classes of
couplings separately and benefit from a great reduction of unknown
parameters. Beyond leading order, however, form factors of any
symmetry can contribute to $E[\obs_i]$:
\begin{itemize}
\item in the integrated cross section and thus in the denominator of
  $E[\obs_i] - E_0[\obs_i]$ in \eqref{expect} couplings of all four
  classes enter quadratically, couplings of class $(a)$ also appear
  linearly;
\item if $\obs_i$ belongs to class $(a)$ the numerator of $E[\obs_i] -
  E_0[\obs_i]$ has terms linear in the couplings of this class but
  couplings of all four classes enter quadratically through
  $q_{ijk}\,$;
\item if $\obs_i$ belongs to a $\CP$ ($\CPT$) odd coupling then the
  numerator in \eqref{expect} is only linear in $\CP$ ($\CPT$) odd
  couplings, but it contains also quadratic terms where a $\CP$
  ($\CPT$) odd coupling is multiplied with a $\CP$ ($\CPT$) even one.
\end{itemize}
We remark that this leads to different behaviours of $E[\obs_i]$ as
one or more couplings $g_i$ become large: whereas for observables in
classes $(b)$, $(c)$ and $(d)$ the expectation value goes to zero when
a coupling of the same class goes to plus or minus infinity the
corresponding limit of an observable in class $(a)$ can be a positive
or negative constant or zero.

In a nonlinear analysis one will therefore in principle have to
consider couplings with all symmetries at the same time. In practice
one might choose simpler procedures if the linear approximation is
expected to be not too bad and if one wants to calculate corrections
to it. One might for instance first analyse the four symmetry classes
separately, neglecting in each case the contributions of the three
other classes at the r.h.s.\ of \eqref{mean} and then refine the
analysis of a class by taking the values obtained in the first step
for the couplings in the other classes as fixed in \eqref{mean}.

We emphasise that even beyond the leading order approximation it is
still true that a nonzero mean value of a $\CP$ or a $\CPT$ odd
observable is an unambiguous sign of $\CP$ violation or the presence
of absorptive parts in the process, respectively. The extraction of
the values of the couplings, however, becomes more involved than in
leading order.

\section{Diagonalisation in the couplings}
\label{sec:diagon}

We shall now propose a method to analyse the data which presents
several advantages in view of the basic problem posed by the large
number of unknown three-boson couplings: with limited event statistics
significant error bounds can only be obtained for subsets of the
coupling parameters, but imposing constraints on the couplings to
reduce their number entails a loss of information that cannot be
retrieved. In view of this it should be advantageous to use a
parametrisation of the couplings which in a given process and at a
given c.m.\ energy has the following properties:
\begin{enumerate}
\item It allows to find
  observables which are only sensitive to one particular coupling
  parameter.
\item The induced errors on the couplings
  determined from these observables are statistically independent.
\end{enumerate}
With this we can on the one hand give single errors for each parameter, on
the other hand we can recover from the single errors the
multidimensional error of the full set of couplings, having avoided
the loss of information incurred by imposing constraints. From the
single errors we can also directly see which combinations of
couplings in more conventional parametrisations can be measured with
good accuracy and to which one is rather insensitive.

\begin{figure}
  \begin{center}
    \leavevmode
    \begin{picture}(0,0)%
      \epsfig{file=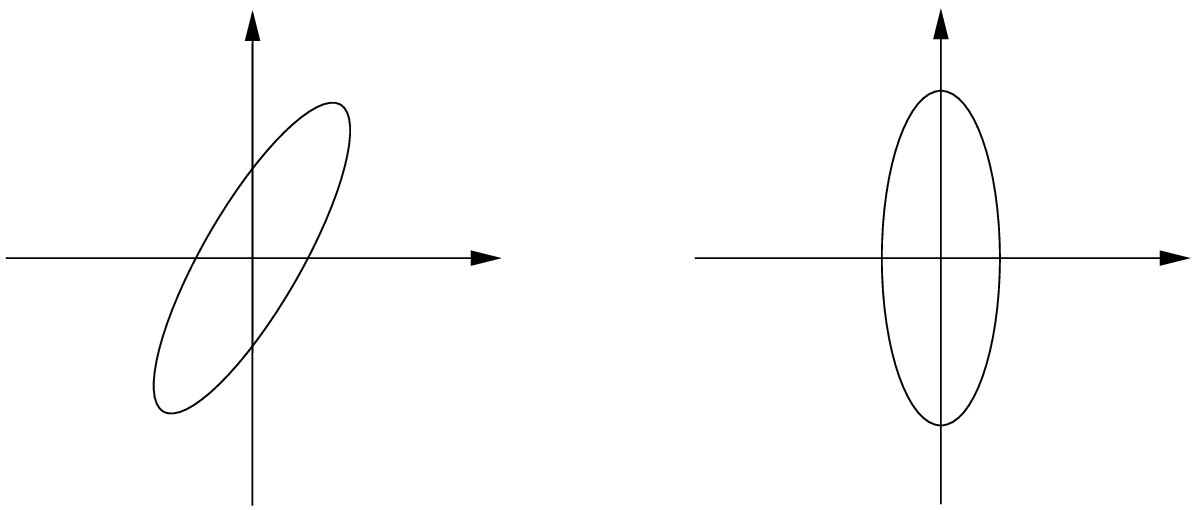}%
    \end{picture}%
    \setlength{\unitlength}{0.00087500in}%
    \begin{picture}(5502,3031)(889,-4805)
      \put(1981,-2401){\makebox(0,0)[lb]{\smash{$g_j$}}}
      \put(5131,-2401){\makebox(0,0)[lb]{\smash{$g_j'$}}}
      \put(6391,-3706){\makebox(0,0)[lb]{\smash{$g_i'$}}}
      \put(3241,-3706){\makebox(0,0)[lb]{\smash{$g_i$}}}
      \put(1116,-2006){\makebox(0,0)[lb]{\smash{\large $(a)$}}}
      \put(4176,-2006){\makebox(0,0)[lb]{\smash{\large $(b)$}}}
    \end{picture}
    \caption{{}Example of the 1--$\sigma$ ellipsis for
      two extracted parameters in a parametrisation where their errors
      are correlated $(a)$ or uncorrelated $(b)$. The single errors on
      the couplings are given by the projection of the ellipses on the
      coordinate axes.}
    \label{fig:correlations}
  \end{center}
\end{figure}

Let us remark that in the leading order analysis there is a set of
observables satisfying point 1.\ in \emph{any} parametrisation of the
couplings. The linear combinations
\begin{equation}
  \label{obsModified}
  {\cal C}_i = \sum_j c^{-1}{}_{i j} \, \obs_j
\end{equation}
of our optimal observables \eqref{optimal} are only sensitive to $g_i$
for each $i$ (cf.\ also \cite{Grzad}). The errors on the couplings
determined from these observables are, however, in general not
uncorrelated; in fact their correlations are the same as those
obtained with the original set $\obs_i$. This can be seen as follows:
going from the $\obs_i$ to the ${\cal C}_i$ we must replace
\begin{eqnarray}
  \label{replaceMatrices}
  c_{ij} &\to& \delta_{ij}  \eqcm \nonumber \\
  V(\obs) &\to& V({\cal C}) = c^{-1} \cdot V(\obs) \cdot (c^{-1})^{T}
  \eqcm
\end{eqnarray}
so that we have from \eqref{couplingsCov}
\begin{equation}
  \label{sameCorrel}
  \left. V(g)^{-1} \right|_{\obs} = N \, c^T \cdot V(\obs)^{-1} \cdot
  c = N \, V({\cal C})^{-1} = \left. V(g)^{-1} \right|_{{\cal C}}
  \eqpt
\end{equation}
In such a case the single errors give an incomplete picture of the
situation if correlations are large. This is illustrated in
fig.~\ref{fig:correlations} $(a)$, where the 1--$\sigma$ ellipsis for
two parameters is shown. Their single errors are given by its
projection on the coordinate axes and in our example are both rather
large. Some linear combinations of them are however measurable with
much better precision, which one can only recognise if both errors and
their correlations are given. In fig.~\ref{fig:correlations} $(b)$
where a set of couplings leading to uncorrelated errors is used the
situation is much simpler. Note also that the number of correlations,
i.e.\ off-diagonals in $V(g)$, is yet modest for two couplings but
increases rapidly with their number.

We will now first see that a parametrisation of the couplings
satisfying both points 1.\ and 2.\ above can be found in idealised
circumstances, and then mention the restrictions one will encounter
under more realistic assumptions.

If the leading order analysis is a sufficiently good approximation the
solution to our problem is easily found. Starting from a set of
couplings $g_i$ and the corresponding optimal observables $\obs_i$ in
\eqref{optimal} we can go to another set $g_i'$ by
\begin{equation}
  \label{couplingsTransf}
  \bvec{g}' = A^{-1}  \bvec{g}  \eqcm
\end{equation}
where we use vector and matrix notation. The coefficients in the
expansion of the differential cross section and the optimal
observables transform as follows:
\begin{eqnarray} \label{obsTransf}   
  \bvec{S}_1' = A^{T} \bvec{S}_1  \nonumber \\
  \bvec{\obs}' = A^{T} \bvec{\obs}  
\end{eqnarray}
Let now $\obs_i$ be an arbitrary set of observables related to $g_i$
and define the corresponding $\obs_i'$ related to $g_i'$ as in
(\ref{obsTransf}). Then we have for the matrices relevant for our
analysis the following transformation properties:
\begin{eqnarray} \label{matrixTransf}
  c' &=& A^{T} \cdot c \cdot A  \nonumber \\
  V(\obs') &=& A^{T} \cdot V(\obs) \cdot A  \nonumber \\
  V(g')^{-1} &=& A^{T} \cdot V(g)^{-1} \cdot A  \eqpt
\end{eqnarray}
As shown in \cite{OldObs} our optimal observables satisfy $V(\obs) =
c$ and $V(g)^{-1} = N c$ so that for them one can choose a
transformation $A$ which diagonalises all three matrices. This new set
$g'$ of parameters obviously has the properties 1.\ and 2.\ we were
looking for.

Beyond the linear approximation of \eqref{expect} the expectation
value of $\obs_i'$ will still receive contributions from several
couplings. In fact there is no set of observables for which the full
nonlinear expression in \eqref{expect} satisfies point 1.\ exactly,
because the denominator involves quadratic terms in \emph{all}
couplings, and this cannot be changed by any linear transformation of
the couplings. If on the other hand the statistical errors are too
large the covariance matrix $V(g)$ will not give a good picture of the
errors as we discussed in sec.~\ref{sec:method}, and its
diagonalisation will not ensure point 2. In the case however where
nonlinear effects in the determination of the couplings and their
errors are not too large, i.e.\ where the leading order expressions
are a good first approximation both points 1.\ and 2.\ above will
still be \emph{approximately} satisfied in a full nonlinear analysis.

We remark that if one has some previous estimates $\tilde{g}_i$ of the
couplings that considerably deviate from zero one may reduce nonlinear
effects in the determination of the $g_i$ by working with an expansion
of $d\sigma / d\phi$ around the $\tilde{g}_i$ as shown in
sec.~\ref{sec:method}; in our diagonalisation programme one will then
use couplings $g_i - \tilde{g}_i$ instead of $g_i$, the matrix
$\tilde{c}$ instead of $c$ etc.

To the extent that the observables \eqref{optimal} are constructed
from expressions of $S_0$ and $S_{1,j}$ which are only approximations
of those that determine the experimentally observed kinematical
distribution the matrices $c$, $V(\obs)$ and $N^{-1} V(g)^{-1}$ will
not quite be the same and cannot be diagonalised at the same time. One
can then diagonalise either $V(\obs)$ or $V(g)^{-1}$ because they are
by definition symmetric and positive definite, whereas $c$ is not
necessarily so.  Again, unless such effects are large one will end up
with a matrix $c'$ that is not diagonal but has relatively small
off-diagonals.

It should also be borne in mind that the covariance matrix $V(g)$ only
gives the statistical errors on the couplings, so that even if it is
exactly diagonal the final errors may be correlated due to
systematics.

\subsection{Simultaneous diagonalisation of the correlation \protect\newline
  matrix and the quadratic term in the total cross section}
\label{sec:simultan}

\begin{figure}[h]
  \begin{center}
    \leavevmode
    \begin{picture}(0,0)%
      \epsfig{file=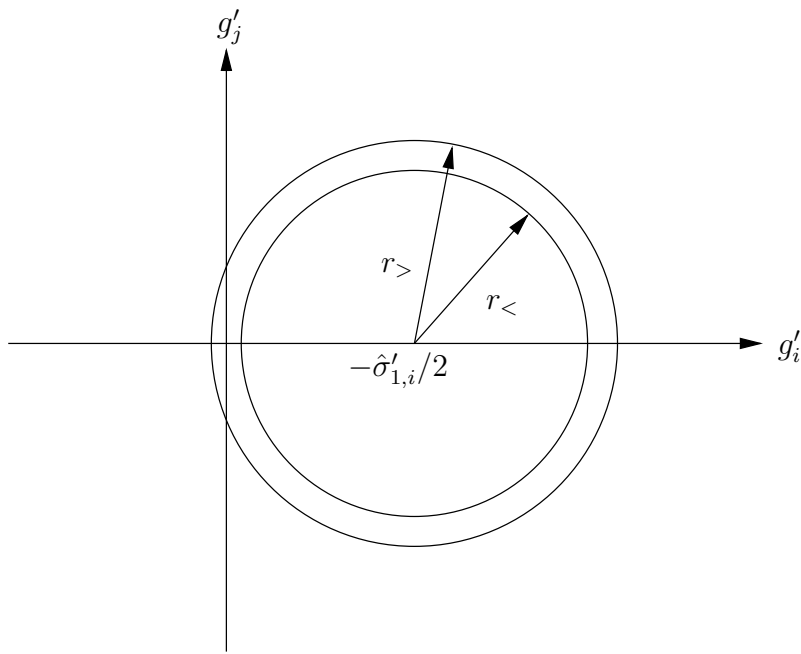}%
    \end{picture}%
    \setlength{\unitlength}{0.00087500in}%
    \begin{picture}(4647,3843)(1339,-5473)
      \put(4231,-3391){\makebox(0,0)[lb]{\smash{$r_{<}$}}}
      \put(2611,-1726){\makebox(0,0)[lb]{\smash{$g_j'$}}}
      \put(5986,-3661){\makebox(0,0)[lb]{\smash{$g_i'$}}}
      \put(3401,-3796){\makebox(0,0)[lb]{\smash{$-\sig{1,i}' /2$}}}
      \put(3601,-3166){\makebox(0,0)[lb]{\smash{$r_{>}$}}}
    \end{picture}
    \caption{{}The measurement of the integrated cross section
      constrains the couplings to a shell between two hyperspheres
      with radii $r_{>}$ and $r_{<}$ given by \protect\eqref{radii}.
      $g_i'$ belongs to symmetry class $(a)$ introduced in
      sec.~\protect\ref{sec:symmetries} and $g_j'$ to class $(b)$,
      $(c)$ or $(d)$. In the example shown here the measurement is
      compatible with both couplings being zero.}
    \label{fig:shell}
  \end{center}
\end{figure}

The choice of transformation in \eqref{couplingsTransf} to
\eqref{matrixTransf} is not unique if one does not require $A$ to be
orthogonal.\footnote{Orthogonal transformations have been used in the
  second paper of ref.~\cite{Papadop}.} We see in fact no strong
argument in favour of an orthogonal transformation and remark that the
various parametrisations of the $WW\gamma$ and $WWZ$ couplings in the
literature are related by non-orthogonal linear transformations. The
freedom to choose $A$ can be used to impose additional conditions on
the transformation, and the one we propose here is that the
transformed quadratic coefficient
\begin{equation}
  \label{quadratTransf}
  \sig{2}' = A^{T} \cdot \sig{2} \cdot A
\end{equation}
in the integrated cross section be the unit matrix. In terms of the
new couplings one then has
\begin{equation}
  \label{XsecTransf}
  \sigma / \sigma_0 = 1 + \sum_{i=1}^{8} \sig{1,i}' \, g_i' +
  \sum_{i=1}^{28} \left( g_i' \right)^2  \eqcm
\end{equation}
where we choose the numbering such that $g_1'$ to $g_8'$ belong to
symmetry class $(a)$ introduced in sec.~\ref{sec:symmetries}, i.e.\ 
they are the $\CP$ and $\CPT$ even couplings. Only these appear
linearly in the cross section, whereas all couplings give a quadratic
contribution with coefficient one.  Having $\sig{2,ij}' = \delta_{ij}$
leads to a convenient simplification of \eqref{expect}, \eqref{mean}.
Moreover, the measurement of the total cross section gives
complementary information on the unknown couplings. Rewriting
\eqref{XsecTransf} as
\begin{equation}
  \label{completSquare}
  \sigma / \sigma_0 = 1 - \sum_{i=1}^{8} \frac{(\sig{1,i}')^2}{4} +
  \sum_{i=1}^{8} \left( g_i' + \frac{\sig{1,i}'}{2} \right)^2 +
  \sum_{i=9}^{28} (g_i')^2
\end{equation}
we see that measuring a cross section $\sigma_{\it exp}$ within an
error $\Delta \sigma$ constrains the couplings to be in a shell
between two hyperspheres with centre at
\begin{equation}
  \label{centre}
  g'_i = - \frac{\sig{1,i}'}{2}  \hspace{1em}  (i=1, \ldots ,8) \eqcm
  \hspace{2em}
  g'_i = 0  \hspace{1em}  (i=9, \ldots ,28)
\end{equation}
in the space of all couplings as shown in fig.~\ref{fig:shell}. Their
radii are given by
\begin{equation}
  \label{radii}
  r_{>}^2 = \frac{\sigma_{\it exp} + \Delta \sigma - \sigma_{\it
      min}}{\sigma_0}
  \eqcm \hspace{2em}
  r_{<}^2 = \frac{\sigma_{\it exp} - \Delta \sigma - \sigma_{\it
      min}}{\sigma_0}  \eqpt
\end{equation}
Here
\begin{equation}
  \label{minXsec}
  \sigma_{\it min} = \sigma_0 \left[ 1 - \sum_{i=1}^{8}
    \frac{(\sig{1,i}')^2}{4} \right]
\end{equation}
is the smallest value the cross section can attain; that such a
minimum exists has been pointed out in \cite{Gaemers}. If in
\eqref{radii} $r_{>}^2$ is positive but $r_{<}^2$ negative the
couplings are inside the hypersphere with radius $r_{>}$, and if both
$r_{>}^2$ and $r_{<}^2$ are negative the ansatz \eqref{diffXsection}
for the cross section is inconsistent with the data within the error
$\Delta \sigma$.

Such constraints can be useful to find the physical set of couplings
when the solution of \eqref{mean} from the measurement of the optimal
observables is not unique. If they are strong enough they might even
restrict the couplings to the region where \eqref{mean} can be
linearised and thus simplify their extraction. One can of course use
the information from the integrated cross section working with any set
of couplings, but again the situation is particularly simple with the
form \eqref{XsecTransf}.

We note that the information from the total rate is complementary to
what is extracted from the mean values of our observables, which
involve normalised kinematical distributions. From an experimental
point of view their respective measurements will presumably have quite
different systematic errors. Let us also recall that the the
measurement of the mean values $\mean_i$ times the number $N$ of
events obtained with a fixed integrated luminosity combines the
information of both \cite{OldObs}. A nonlinear data analysis as
presented in sec.~\ref{sec:method} can also be done in this case. We
draw attention to the fact that unphysical solutions of
equation~\eqref{mean} for $\mean_i$ and of its analogue for $N
\mean_i$ will in general be different. We shall however not elaborate
on this point here.

Another aspect of the couplings with the property \eqref{XsecTransf}
is the following. It is well known that constant coupling parameters
deviating from the standard model tree level values lead to amplitudes
that violate unitarity \cite{unitarity}.  The coefficients $\sig{1,i}$
and $\sig{2,ij}$ in the total cross section $\sigma$ increase strongly
with the $e^+ e^-$ c.m.\ energy $\sqrt{s}$ and the couplings $g_i$
must vanish as $s$ becomes large to ensure a decent high-energy
behaviour of $\sigma$. In our new parametrisation the quadratic
coefficients in $\hat\sigma_i$ are energy independent, and in this
sense the new couplings are at a ``natural scale'' at every energy.

To complete this section we show that a transformation with the
properties we require always exists, i.e.\ that we can find a matrix
$A$ that diagonalises $V(g)^{-1}$ in \eqref{matrixTransf} and
transforms $\sig{2}$ in \eqref{quadratTransf} to the unit matrix. The
argument is analogous if one replaces $V(g)^{-1}$ with $V(\obs)$. By
construction both $V(g)^{-1}$ and $\sig{2}$ are symmetric and positive
definite, so our problem is the same as finding normal coordinates for
a multidimensional harmonic oscillator in classical mechanics (cf.\ 
e.g.~\cite{Gold}). To make this analogy transparent let us write
${\cal T} = \sig{2}$ and ${\cal V} = V(g)^{-1}$; we then have to find
$A$ so that
\begin{equation}
  \label{diagonalise}
  A^{T} \cdot {\cal T} \cdot A = 1 \eqcm \hspace{4em}
  A^{T} \cdot {\cal V} \cdot A = D
\end{equation}
with $D$ being diagonal. The elements $d_{i}$ of $D$ are generalised
eigenvalues of ${\cal V}$ satisfying
\begin{equation}
  \label{eigenvalues}
  {\cal V} \, \bvec{a}_i = d_i \: {\cal T} \bvec{a}_i  \eqcm
\end{equation}
where $\bvec{a}_i$ is the $i$-th column vector of $A$. The solution is
well known to be
\begin{equation}
  \label{solution}
  A = {\cal T}^{-1/2} \cdot U
\end{equation}
where $U$ is the orthogonal matrix that transforms ${\cal T}^{-1/2}
\cdot {\cal V} \cdot {\cal T}^{-1/2}$ to $D$. Of course one need not
use \eqref{solution} in practice as there are convenient algorithms
available to find $A$ and $D$. In our numerical calculations we have
used the routine \texttt{Eigenvals} of the algebraic package MAPLE.

\section{Numerical examples}
\label{sec:numeric}

We will now give some numerical examples of our method described in
the previous section. In this section we will stay within the
framework of a leading order analysis of the observables. We start
from the results in \cite{OldObs}, where the sensitivity of
optimal observables for semileptonic $WW$-decays was calculated. We
assume a full kinematical reconstruction of the final state, except
for the ambiguity one is left with if the jet charge is not known. For
the standard model cross section we use the Born approximation and
neglect effects of the finite $W$ width. To describe the triple boson
couplings we take the form factors $f_i^\gamma$, $f_i^Z$ ($i = 1
\ldots 7$) of \cite{Hagiwa}; deviations from their standard model tree
level values will be referred to as ``anomalous couplings''.

Let us first look at a c.m.\ energy of $\sqrt{s} = 190 \GeV$, which
will be attained at LEP2. The coefficient matrix $c$ can be found in
table~4 of \cite{OldObs} and in table~\ref{tab:coeff190} here we give
the diagonal elements of the transformed matrix $c'$, ordered
according to the symmetry of the corresponding observables.  The one
standard deviation ellipsoid is diagonal in the couplings $g_i'$, thus
its intersections with the coordinate axes equal its projections on
these axes. The errors $\delta g_i'$ setting all other $g_j'$ to zero
are then equal to the errors $\Delta g_i'$ where all other $g_j'$ are
arbitrary. They are given by
\begin{equation}
  \label{oneSigma}
  \Delta g_i' = \frac{1}{\sqrt{V(g')^{-1}{}_{ii}}}
  = \frac{1}{\sqrt{N c_{ii}'}}
\end{equation}
and are listed in table~\ref{tab:sens190} for an integrated luminosity
of $500 \pbarn^{-1}$.

We immediately remark that a negative diagonal element occurs in the
transformed coefficient matrix, which is not allowed because $c' =
V(\obs')$ is a covariance matrix and thus positive definite. We
encounter here a problem of numerical instability: small errors in the
calculation of the original matrices $c$ and $\sig{2}$ can have a
large effect on the smallest generalised eigenvalues $c_{ii}'$ and
their eigenvectors, even to the point that eigenvalues come out with
the wrong sign. This is not only a problem of our particular way of
diagonalisation, but also occurs if one diagonalises $c$ with an
orthogonal matrix; we find that one of the usual eigenvalues of $c$ in
the subspace of couplings with symmetry $(b)$ is negative. Such
instabilities can cause large errors in the matrix inversion of $c$
and $V(\obs)$. One needs $V(\obs)^{-1}$ to calculate the error on the
extracted couplings as can be seen from \eqref{chi} and
\eqref{couplingsCov}, and large errors on $c^{-1}$ can lead to large
uncertainties in the extracted couplings, irrespective of whether
$c^{-1}$ is explicitly used to solve the system \eqref{mean}. One will
of course aim to calculate $c$ and $V(\obs)$ with best possible
precision, but such an effort has limits, in particular if they are
determined from simulated events and include for instance radiative
corrections or detector effects. On a more fundamental level any
calculation of these matrices will only be an approximation of the
``exact'' ones that correspond to the kinematical distributions seen
in experiment. In this sense it seems quite inevitable that small
eigenvalues (the usual or our generalised ones) of $c$ and $V(\obs)$
and their eigenvectors are sensitive to imprecisions in the
calculation and can lead to large errors or uncertainties in the data
analysis.  This holds of course even if one does not obtain
eigenvalues with the wrong sign. We think that also in view of this a
diagonalisation is useful, not because it solves the problem but
because it makes it explicit! It allows to easily identify those
combinations of couplings which have small corresponding eigenvalues
in $c$ and $V(\obs)$ and will be the most unsafe ones in the analysis.
From \eqref{oneSigma} we see that they are those combinations for
which the statistical errors will be largest. Here the most unsafe
coupling parameter is $g_{16}'$. One might thus choose to exclude it,
and possibly other couplings, from the analysis and work in the
remaining subspace of the $g_i'$ where the numerics is more stable and
where in any case the experiment is most sensitive. We will come back
to this in sec.~\ref{sec:practice}.

\begin{table}
  \begin{center}
    \leavevmode
    \caption{\label{tab:coeff190}Diagonal elements $c_{ii}'$ of the
      transformed coefficient matrix at $\protect\sqrt{s} = 190 \GeV$.
      They are ordered by the symmetry classes $(a)$ to $(d)$
      introduced in sec.~\protect\ref{sec:symmetries}; the first row
      contains $c_{1,1}'$ to $c_{8,8}'$, the second $c_{9,9}'$ to
      $c_{16,16}'$ etc. Note that the $c_{ii}'$ are by construction
      positive; the negative value for $c_{16,16}'$ is due to
      numerical integration errors in the original matrix $c$. This
      problem is further discussed in the text. A similar comment
      applies to tables \protect\ref{tab:lr190},
      \protect\ref{tab:coeff500} and \protect\ref{tab:lr500}.}
    \vspace{\baselineskip}
    \begin{tabular}{crrrrrrrr}
      \hline
      $(a)$ & 1.5  & 1.4 & 0.83 & 0.70 & 0.32  & 0.10   & 0.027  &
      0.019 \\
      $(b)$ & 1.1  & 1.0 & 0.81 & 0.68 & 0.028 & 0.0093 & 0.0056 &
      $- 0.00045$ \\
      $(c)$ & 0.80 & 0.67 & 0.28 & 0.028 & 0.013 & 0.012 \\
      $(d)$ & 1.4  & 1.2  & 0.85 & 0.13  & 0.039 & 0.017 \\
      \hline
    \end{tabular}
  \end{center}
  \vspace{\baselineskip}
  \begin{center}
    \leavevmode
    \caption{\label{tab:sens190}1--$\sigma$ errors $\Delta g_i'$ on
      the extracted couplings corresponding to the coefficients
      $c_{ii}'$ in table~\protect\ref{tab:coeff190}. They are
      calculated from \protect\eqref{oneSigma} for an integrated
      luminosity of $500 \pbarn^{-1}$.}
    \vspace{\baselineskip}
    \begin{tabular}{crrrrrrrr}
      \hline
      $(a)$ & 0.015 & 0.016 & 0.020 & 0.022 & 0.032 & 0.058 & 0.11 &
      0.13 \\
      $(b)$ & 0.017 & 0.018 & 0.020 & 0.022 & 0.11  & 0.19  & 0.24 &
      \multicolumn{1}{c}{---} \\
      $(c)$ & 0.020 & 0.022 & 0.035 & 0.11  & 0.16  & 0.17 \\
      $(d)$ & 0.015 & 0.017 & 0.020 & 0.051 & 0.093 & 0.14 \\
      \hline
    \end{tabular}
  \end{center}
  \vspace{\baselineskip}
  \begin{center}
    \leavevmode
    \caption{\label{tab:lr190}Diagonal elements $c_{ii}'$ of the
      coefficient matrix restricted to the left or right handed
      subspace of the couplings as explained in the text. The values
      in the left handed subspace differ from the corresponding ones
      in table 1 by at most 3\%.}
    \vspace{\baselineskip}
    \begin{tabular}{crrrrrrrrr}
      & \multicolumn{4}{c}{left handed} & \hspace{1em} &
      \multicolumn{4}{c}{right handed} \\
      \cline{1-5} \cline{7-10}
      $(a)$ & 1.5 & 1.4 & 0.83 & 0.69 & & 0.35  & 0.12 & 0.030 & 0.019
      \\
      $(b)$ & 1.1 & 1.0 & 0.81 & 0.68 & & 0.030 & 0.23 & 0.11  &
      $- 0.00042$ \\
      $(c)$ & 0.80 & 0.67 & 0.28 & & & 0.030 & 0.018 & 0.012 & \\
      $(d)$ & 1.4  & 1.2  & 0.85 & & & 0.13  & 0.042 & 0.018 & \\
      \cline{1-5} \cline{7-10}
    \end{tabular}
  \end{center}
\end{table}

In tables~\ref{tab:coeff190} and \ref{tab:sens190} we find that the
range of sensitivities is quite large, typically spanning several
orders of magnitude in the $c_{ii}'$. We can actually identify the
role of those form factors whose $c_{ii}'$ are small. To this end we
pass from the usual form factors $f_i^{\gamma}$, $f_i^Z$ to the
combinations which appear in the amplitudes for $e^+ e^- \to W^+ W^-$
with left or right handed electron polarisation, respectively:
\begin{eqnarray}
  \label{leftRight}
  f^{L}_{i} & = & 4 \wein \, f^{\gamma}_{i} + (2 - 4 \wein)\, \xi
  f^{Z}_{i},     \nonumber \\
  f^{R}_{i} & = & 4 \wein \, f^{\gamma}_{i} - 4 \wein\, \xi f^{Z}_{i},
\end{eqnarray}
where $\xi = s / (s - M_Z^2)$ and $\theta_W$ is the weak mixing angle.
The matrix $\sig{2}$ is block diagonal in the left and right handed
form factors because different electron polarisations do not interfere
in the cross section, but the coefficient matrix $c$ is not. If one
sets all right (left) handed anomalous couplings to zero then one has
to diagonalise $c$ in the left (right) handed subspace. Doing this we
obtain the results shown in table~\ref{tab:lr190}. In each of the
classes $(a)$ to $(d)$ we find a clear correspondence of the
generalised eigenvalues in the left (right) handed subspace with the
largest (smallest) ones of the full matrix $c$ given in
table~\ref{tab:coeff190}. The form factors $g_i'$ to which one is most
sensitive thus correspond predominantly to left handed combinations,
whereas the right handed combinations are more difficult to measure.
This confirms our findings in \cite{OldObs} and can be explained by
the missing neutrino exchange graph for right handed electrons which
can give a large interference with anomalous triple boson couplings.
As a word of caution we remark that the values given in
table~\ref{tab:lr190} do not correspond to those for left or right
handed electron beams, because for unpolarised beams both electron
helicities contribute to the standard model cross section even if
anomalous couplings corresponding to one helicity are (assumed to be)
zero. We found however in \cite{OldObs} that the difference between
$c$ for a left handed electron beam and the left handed submatrix of
$c$ for unpolarised beams is small, again because the right handed
standard model contribution is small compared to the left handed one
due to the missing neutrino exchange graph.

Let us now come to the integrated cross section. We first give it in
the parametrisation by $f_i^\gamma$ and $f_i^Z$, where the standard
model tree values are $f_1^\gamma = f_1^Z = 1$, $f_3^\gamma = f_3^Z =
2$ and zero for all other form factors:
\begin{eqnarray}
  \label{XsecOld190}
  \lefteqn{\sigma /\sigma_0 = 1} \nonumber \\
  && {} + 0.022 \, (\re f^\gamma_1 - 1) + 0.013 \, (\re f^Z_1 -1) -
  0.031 \, \re f^\gamma_2 - 0.010 \, \re f^Z_2 \nonumber\\
  && {} - 0.074 \, (\re f^\gamma_3 -2) - 0.019 \, (\re f^Z_3 - 2) +
  0.0058 \, \re f^\gamma_5 + 0.0092 \, \re f^Z_5 \nonumber\\
  && {} + \{ \textrm{quadratic terms} \}
\end{eqnarray}
We do not give the full matrix $\sig{2}$ for the quadratic terms here,
but remark that the absolute values of its elements are between 0.001
and 0.3 and its eigenvalues between 0.004 and 0.4. In our new
parametrisation we have
\begin{eqnarray}
  \label{XsecNew190}
  \sigma /\sigma_0 &=&
  1 + 0.18 \, g_1' + 0.16 \, g_2' - 0.0053 \, g_3' - 0.052 \, g_4'
  \nonumber \\
  && \phantom{1} - 0.15 \, g_5' + 0.071 \, g_6' - 0.029 \, g_7' +
  0.0091 \, g_8' + \sum_{i=1}^{28} (g_i')^2
  \nonumber \\
  &=& 1 - 0.023 \nonumber \\
  && {} + (g'_1 + 0.092)^2  + (g'_2 + 0.082)^2  + (g'_3 - 0.0026)^2 +
  (g'_4 - 0.026)^2  \nonumber \\
  && {} + (g'_5 - 0.073)^2  + (g'_6 + 0.035)^2 + (g'_7 - 0.014)^2 +
  (g'_8 + 0.0046)^2 \nonumber \\
  && {} + \sum_{i=9}^{28} (g_i')^2  \eqpt
\end{eqnarray}
Comparing with the first row of table~\ref{tab:coeff190} we see that
couplings whose linear contribution to the cross section is relatively
small can give a relatively large linear contribution to their optimal
observable and vice versa. A measurement with an integrated luminosity
of $500 \pbarn^{-1}$ at $\sqrt{s} = 190 \GeV$ that finds the cross
section equal to its standard model (Born level) value $\sigma_0$ will
have a relative statistical error $\Delta \sigma / \sigma_{\it exp} =
1/ \sqrt{N} = 0.018$. According to \eqref{radii} this measurement
would constrain the couplings $g_i'$ to be between hyperspheres with
radii $r_{<} = 0.066$ and $r_{>} = 0.202$ in their 28-dimensional
space. Comparing with table~\ref{tab:sens190} we see that the
thickness $r_{>} - r_{<} = 0.136$ of this shell is of the same order
of magnitude as the largest statistical errors to be achieved with our
optimal observables, which give the extensions of the 28-dimensional
error ellipsoid in a linearised analysis. This reflects the well-known
fact that the integrated cross section is clearly not as sensitive to
anomalous couplings as the detailed kinematical distributions whose
information we extract with our observables. It gives however an
additional constraint that should provide a means to constrain the
couplings such as $g_{16}'$ which are practically unmeasurable by
normalised observables. Also we should get a useful cross check and
help selecting the physical solution of \eqref{mean} in a full
nonlinear analysis.

Let us now apply our method to a typical LC energy $\sqrt{s} = 500
\GeV$ with an integrated luminosity of $10 \fbarn^{-1}$ (cf.\ 
tables~\ref{tab:coeff500} to \ref{tab:lr500}). Here we encounter the
particularity that the coefficients of the form factors $f_2^{\gamma,
  Z}$ and $f_7^{\gamma, Z}$ in the scattering amplitude grow faster
with energy than those of the other couplings \cite{Hagiwa}. As a
result the off-diagonal matrix elements in $c$ between $f_2^{\gamma,
  Z}$ and another coupling are about a factor of 10 larger than matrix
elements not involving $f_2^{\gamma, Z}$, and elements involving only
$f_2^{\gamma, Z}$ are larger by yet another factor of 10. The (usual)
eigenvalues of $c$ in the $\CP$ conserving sector span 6 orders of
magnitude. In the $\CP$ violating sector the situation is less
dramatic, but still elements of $c$ involving $f_7^{\gamma, Z}$ alone
are about a factor of 10 larger than the others. The same phenomenon
is found in the cross section, which reads
\begin{eqnarray}
  \label{XsecOld500}
  \lefteqn{\sigma / \sigma_0 = 1} \nonumber \\
  && {} + 0.45 \, (\re f^\gamma_1 - 1) + 0.23 \, (\re f^Z_1 -1) - 8.3
  \, \re f^\gamma_2 - 4.3 \, \re f^{Z}_{2} \nonumber\\
  && {} - 0.58 \, (\re f^\gamma_3 -2) - 0.31 \, (\re f^Z_3 - 2) + 0.056
  \, \re f^{\gamma}_{5} + 0.070 \, \re f^{Z}_{5} \nonumber\\
  && {} + \{ \textrm{quadratic terms} \}
\end{eqnarray}
where $\sig{2}$ has elements with absolute values between 0.04 and
1700 and eigenvalues between 0.02 and 1800. After our simultaneous
diagonalisation the range of the matrix elements $c_{ii}'$ is
significantly smaller as can be seen in table~\ref{tab:coeff500}. The
cross section now reads
\begin{eqnarray}
  \label{XsecNew500}
  \sigma / \sigma_0 &=&
  1 - 0.31\,g_1'  + 0.057\,g_2' + 0.0018\,g_3' - 0.014\,g_4' \nonumber
  \\
  && \phantom{1} - 0.026\,g_5' - 0.044\,g_6' - 0.014\,g_7' -
  0.013\,g_8' + \sum_{i=1}^{28} (g_i')^2 \nonumber \\
  &=& 1 - 0.026 \nonumber \\
  && {} + (g_1' - 0.16)^2 + (g_2' + 0.028)^2 + (g_3' +
  0.00091)^2 + (g_4' - 0.0068)^2 \nonumber \\
  && {} + (g_5' - 0.013)^2 + (g_6' - 0.022)^2 +
  (g_7' - 0.0069)^2 + (g_8' - 0.0066) ^2 \nonumber \\
  && {} + \sum_{i=9}^{28} (g_i')^2  \eqpt
\end{eqnarray}
Of course, the $g_i'$ are now in general energy-dependent for constant
form factors $f_i^{\gamma,Z}$.  As we mentioned in
sec.~\ref{sec:simultan} the rapid growth with $s$ of the coefficients
of anomalous parts of the couplings $f_i^{\gamma,Z}$ in the amplitude
has been ``absorbed'' into the coupling parameters by the
transformation of $\sig{2}$ to the unit matrix. As a result the linear
coefficients in the cross section and the elements of the coefficient
matrix $c$ at $\sqrt{s} = 500 \GeV$ have the same order of magnitude
as at LEP2 energies, and the large differences in the $s$-dependence
of the coefficients for different couplings have been evened out.

A measurement at the LC that gives the standard model cross section
constrains the couplings to be between 28-dimensional hyperspheres
with radii $r_{<} = 0.140$ and $r_{>} = 0.182$, taking the relative
statistical error of $0.0067$ on the cross section one would obtain
with an integrated luminosity of $10 \fbarn^{-1}$. The thickness of
the shell, $r_{>} - r_{<} = 0.042$, is again of the order of the
largest statistical errors on the $g_i'$ one can obtain with optimal
observables (cf.\ table~\ref{tab:sens500}). Moreover such a small
statistical error is likely to be small compared with systematic
errors, so that the sensitivity of the integrated cross section will
even be less.

\begin{table}
  \begin{center}
    \leavevmode
    \caption{\label{tab:coeff500}As table~\protect\ref{tab:coeff190},
      but for $\protect\sqrt{s} = 500 \GeV$.}
    \vspace{\baselineskip}
    \begin{tabular}{crrrrrrrr}
      \hline
      $(a)$ & 1.4 & 1.2 & 0.74 & 0.65 & 0.33  & 0.11  & 0.056  & 0.033
      \\
      $(b)$ & 1.3 & 1.0 & 0.79 & 0.29 & 0.097 & 0.056 & 0.0092 &
      $- 0.0013$ \\
      $(c)$ & 1.2 & 0.58 & 0.32 & 0.066 & 0.027 & 0.013 \\
      $(d)$ & 1.4 & 1.0  & 0.83 & 0.22  & 0.033 & 0.025 \\
      \hline
    \end{tabular}
  \end{center}
  \vspace{\baselineskip}
  \begin{center}
    \leavevmode
    \caption{\label{tab:sens500}1--$\sigma$ errors $\Delta g_i'$ on
      the extracted couplings, corresponding to the coefficients
      $c_{ii}'$ in table~\protect\ref{tab:coeff500} for an integrated
      luminosity of $10 \fbarn^{-1}$.}
    \vspace{\baselineskip}
    \begin{tabular}{crrrrrrrr}
      \hline
      $(a)$ & 0.0056 & 0.0062 & 0.0078 & 0.0083 & 0.012 & 0.020 &
      0.029 & 0.037 \\
      $(b)$ & 0.0060 & 0.0066 & 0.0075 & 0.013  & 0.022 & 0.028 &
      0.070 & \multicolumn{1}{c}{---} \\
      $(c)$ & 0.0062 & 0.0088 & 0.012  & 0.026 & 0.041 & 0.059 \\
      $(d)$ & 0.0057 & 0.0067 & 0.0074 & 0.014 & 0.037 & 0.043 \\
      \hline
    \end{tabular}
  \end{center}
  \vspace{\baselineskip}
  \begin{center}
    \leavevmode
    \caption{\label{tab:lr500}As table~\protect\ref{tab:lr190} but for
      $\protect\sqrt{s} = 500 \GeV$, to be compared with
      table~\protect\ref{tab:coeff500}.}
    \vspace{\baselineskip}
    \begin{tabular}{crrrrrrrrr}
      & \multicolumn{4}{c}{left handed} & \hspace{1em}
      & \multicolumn{4}{c}{right handed} \\
      \cline{1-5} \cline{7-10}
      $(a)$ & 1.4 & 1.1 & 0.72 & 0.63 & & 0.50 & 0.17  & 0.062 & 0.044
      \\
      $(b)$ & 1.3 & 1.0 & 0.79 & 0.26 & & 0.11 & 0.083 & 0.023 &
      $- 0.0012$ \\
      $(c)$ & 1.2 & 0.58 & 0.32 & & & 0.076 & 0.031 & 0.013 & \\
      $(d)$ & 1.4 & 1.0  & 0.83 & & & 0.24  & 0.040 & 0.026 & \\
      \cline{1-5} \cline{7-10}
    \end{tabular}
  \end{center}
\end{table}

Finally we remark that like in the case for $\sqrt{s} = 190 \GeV$
those couplings $g_i'$ which give the largest statistical errors
in the optimal observable analysis are
predominantly related to right handed combinations of form factors as
can be seen from the comparison of tables~\ref{tab:coeff500} and
\ref{tab:lr500}.

\section{Simultaneous diagonalisation in practice}
\label{sec:practice}

Let us sketch how our method of simultaneous diagonalisation might be
used in practice.

\begin{enumerate}
\item One first has to choose which matrix to diagonalise
  simultaneously with $\sig{2}$. These matrices need not be the same
  ones to be used in the data analysis itself but may be calculated
  under further approximations. Covariance matrices for the
  observables and extracted couplings can be evaluated for zero $g_i$
  as our entire procedure will only have its desired properties if
  nonlinear effects are not too large. If one uses the same
  approximation of the differential cross section \eqref{diffXsection}
  for the construction of the optimal observables \eqref{optimal} and
  the calculation of $c$, $V(\obs)$ and $N^{-1} V(g)^{-1}$ then the
  latter are all equal and can be diagonalised at the same time.
  Otherwise one has to choose a positive definite symmetric matrix for
  the diagonalisation, i.e.\ one of the covariance matrices. The
  calculation of $V(g)$ or of its inverse from \eqref{couplingsCov}
  involves however a matrix inversion and might suffer from numerical
  instabilities, so presumably the best choice will be $V(\obs)$.
\item In the next step one carries out the simultaneous
  diagonalisation of the chosen matrix and $\sig{2}$ as described in
  sec.~\ref{sec:simultan} and determines the transformation matrix $A$
  in \eqref{couplingsTransf} to \eqref{matrixTransf}. At this point it
  will be useful to test the numerical stability of the transformed
  matrices, for instance by re-calculating them in the new basis of
  couplings or by repeating the diagonalisation procedure with
  slightly modified initial matrices. One might choose to discard some
  of the new couplings $g_i'$ and the corresponding observables from
  the analysis if the corresponding matrix elements are found to be
  instable. This does not mean that one has to set these couplings to
  zero. From the measurement of the total cross section one will
  obtain limits on them, which will also allow to control the
  contribution they can give to the mean values of those observables
  that are kept in the analysis because the matrix $c'$ is not exactly
  diagonal and because of nonlinear terms in \eqref{mean}.
\item In the new parametrisation of the couplings one then carries out
  the analysis of the data. Here $V(\obs')$, $c'$, $\sig{2}'$ and the
  other coefficients in \eqref{expect} will be determined under the
  most realistic assumptions and with the best precision one can
  afford. They will not be exactly diagonal in practice, but should
  have small off-diagonal elements if the approximations made in step
  1.\ and in the construction of the optimal observables are
  sufficiently good.
\item One can then give both single and multidimensional errors on the
  measured coupling parameters $g_i'$. At this stage one can also
  present the results in other, more conventional parametrisations of
  the couplings and in particular compare with the measurements of
  other experiments, restricting oneself to whatever subspace of
  couplings might have been chosen there.
\end{enumerate}

\section{Summary}
\label{sec:sum}

In the first part of this paper we have shown how to extract coupling
parameters from the measured mean values $\mean_i$ of appropriate
observables without the approximation that the couplings are small.
Errors on the couplings can be obtained from a $\chi^2$-fit of the
$\mean_i$. If one puts constraints on the couplings in order to reduce
their number the method also gives an indication of how compatible
these constraints are with the data. The ``optimal observables''
discussed in \cite{OldObs} have statistical errors equal to the
smallest possible ones to leading order in the coupling parameters
$g_i$.  Beyond the leading order approximation one can obtain more
sensitive observables if one has some previous estimate $\tilde{g}_i$
for the couplings, expanding the differential cross section around
$\tilde{g}_i$ instead of zero and constructing observables from the
corresponding expansion coefficients. In the appendix we show that up
to linear reparametrisations the choice of optimal observables is
unique: any other set of observables must give bigger (statistical)
errors.

In a second part we have proposed to perform the data analysis using a
particular parametrisation $g_i'$ of the couplings, which is specific
to the process and its c.m.\ energy. It is obtained from the initial
set $g_i$ by a linear transformation which diagonalises the covariance
matrix $V(\obs)$ of the observables and transforms the matrix
$\sig{2}$ of quadratic coefficients in the integrated cross section
\eqref{intXsection} to unity. In an idealised framework each optimal
observable $\obs_i'$ for this parametrisation is only sensitive to one
coupling, and the statistical errors on the extracted couplings are
uncorrelated. Under realistic circumstances both properties can be
approximately satisfied provided that the analysis stays in a region
of parameter space where the dependence of the mean values $\mean_i'$
on the couplings is not far from linear. Various matrices are then
approximately diagonal which should generally facilitate the data
analysis. In particular one can directly give errors on single or a
small number of couplings, which will be necessary to obtain
statistically significant results with a limited number of events. At
the same time one can readily present multidimensional errors in
parameter space, which is essential to compare with the results of
measurements that impose various different constraints on the
couplings. Having approximately diagonal matrices also allows to
easily identify those directions in parameter space which can be
measured best and those for which the statistical errors will be large
and which are likely to be associated with numerical instabilities,
for example in matrix inversions. One can thus recognise and seek to
remedy such problems in an early stage of the analysis.

The measurement of the total cross section $\sigma$ gives valuable
complementary information on the coupling parameters. Its dependence
on the couplings is particularly simple in the parametrisation we
propose since the quadratic contributions are $(g_i')^2$ times the
standard model cross section $\sigma_0$, i.e.\ they have the same form
for all couplings. A measurement of $\sigma$ will then restrict the
$g_i'$ to a shell between two hyperspheres in parameter space.

We have given some numerical examples of our method applied to the
semileptonic decay channels in $e^+ e^- \to W^+ W^-$. In particular we
find that the couplings $g_i'$ which can be measured best with
unpolarised beams predominantly appear in the amplitude for left
handed electrons (or right handed positrons), and that the $g_i'$ with
the largest statistical errors mainly correspond to the opposite
lepton helicity. Comparing our results at LEP2 and LC energies we see
that the coefficients in the linear contributions of the couplings
$g_i'$ to our observables and to the integrated cross section change
much less with energy than in usual parametrisations.  This is because
in the new parametrisation the quadratic coefficients in the
normalised cross section $\sigma / \sigma_0$ are by construction
energy independent.

\section*{Acknowledgements}

We would like to thank Ch.~Hartmann and M.~Kocian for their continued
interest in optimal observables for triple gauge couplings. We
gratefully acknowledge discussions with and remarks by J.~Bl\"umlein,
P.~Overmann, N.~Wermes, \mbox{M.~Wunsch}, and P.~M.~Zerwas.

This work has in part been financially supported by the EU Programme
``Human Capital and Mobility'', Network ``Physics at High Energy
Colliders'', contracts CHRX-CT93-0357 (DG 12 COMA) and
ERBCHBI-CT94-1342, and by BMBF, grant \mbox{05 7HD 91P(0)}. It was
started while one of us (MD) was at the University of Cambridge, and
we acknowledge support by the ARC Programme of the British Council and
the German Academic Exchange Service (DAAD), grant
313-ARC-VIII-VO/scu, which made mutual visits of the Cambridge and
Heidelberg groups possible.

\appendix

\section*{Appendix: Uniqueness of optimal observables}
\label{sec:unique}

In this appendix we show that the set of observables
\eqref{newOptimal}, obtained from expanding the differential cross
section about the actual values of the couplings, is unique in the
sense that up to the linear reparametrisations \eqref{reparam} it is
the only set of $n$ integrated observables which in the limit of large
$N$ leads to the minimum error on the $n$ extracted parameters.

To keep our notation simple we give the proof for the case that the
actual values of the $g_i$ are zero. The expectation value and
covariance of functions $f(\phi)$ and $g(\phi)$ are then given by
\begin{equation}
  \label{zeroOrder}
  E_0[f] = \frac{\int d\phi \, f(\phi)  S_0(\phi)}{\int d\phi \,
    S_0(\phi)}  \eqcm \hspace{3em}
  V_0[f, g] = E_0[f g] - E_0[f] \, E_0[g]  \eqpt
\end{equation}
In the general case one has instead of $S_0$ the zeroth order
coefficient $\widetilde{S}_0$ \eqref{newDiffXsection} from the
expansion about the appropriate values $\tilde{g}_i$.

For large $N$ the covariance matrix for the extracted couplings is
given by \eqref{couplingsCov}. Under a linear
reparametrisation of observables,
\begin{equation}
  \label{reparam}
  \obs_i(\phi) \to \obs_i'(\phi) = \sum_j a_{ij} \obs_j(\phi) + b_i
  \eqcm
\end{equation}
where $a_{ij}$ and $b_i$ are constants and the matrix $a_{ij}$ is
nonsingular, the matrices $c$ from \eqref{obsCoeffCompact} and
$V(\obs)$ transform according to
\begin{eqnarray}
  \label{transformLinear}
  c' &=& a \cdot c  \eqcm \nonumber \\
  V(\obs') &=& a \cdot V(\obs) \cdot a^{T}  \eqpt
\end{eqnarray}
From \eqref{couplingsCov} we see that the covariance matrix $V(g)$ is
unchanged under such a transformation. For our proof we can hence
restrict ourselves to observables with mean value
\begin{equation}
  \label{centred}
  E_0[\obs_i] = 0
\end{equation}
and with a coefficient matrix $c_{ij} = \delta_{ij}$. From
\eqref{obsCoeffCompact} we then have the condition
\begin{equation}
  \label{normalised}
  V_0[\obs_i \, , \; S_{1,j} /S_0] = \delta_{ij}
\end{equation}
and the error on the extracted couplings is given by
\begin{equation}
  \label{simpleCov}
  V(g)_{ij} = N^{-1} V(\obs)_{ij} = N^{-1} V_0[\obs_i, \obs_j]  \eqpt
\end{equation}

From \cite{OldObs} we know that the optimal observables
\eqref{optimal} lead to the smallest possible error on the $g_i$,
given by the Cram\'{e}r-Rao bound. To satisfy our conditions
\eqref{centred} and \eqref{normalised} we take the linear combinations
\begin{equation}
  \label{modifiedAgain}
  {\cal D}_i(\phi) = \sum_j (c_{\it opt})^{-1}{}_{i j} \,
  \left( \rule{0pt}{2ex} S_{1,j}(\phi) / S_0(\phi) - E_0[S_{1,j} / S_0]
  \right)
\end{equation}
with
\begin{equation}
  \label{coefficientAgain}
  (c_{\it opt})_{ij} = V_0[S_{1,i} / S_0 \, , \; S_{1,j} / S_0]  \eqpt
\end{equation}
We assume that the functions $S_{1,i}$ are linearly independent.
Otherwise some of the parameters $g_i$ are superfluous and can be
eliminated; our assumption is thus that the $g_i$ are an independent
set of parameters for the anomalous couplings. Linear independence of
the $S_{1,i}(\phi)$ guarantees that $c_{\it opt}$ is nonsingular,
which has tacitly been used at several instances in our paper. The set
${\cal D}_i$ is related to the optimal observables $S_{1,i} / S_0$ by
a linear transformation \eqref{reparam} and thus gives the same
optimal error matrix $V(g)$.

The covariance $V_0[f,g]$ defines a scalar product on the Hilbert
space ${\cal H}$ of sufficiently smooth functions of $\phi$ with the
property $E_0[f] = 0$.\footnote{A similar scalar product has also been
  used in \cite{Atwood}.} The functions ${\cal D}_i$ span a subspace
${\cal H}^{\it I}$ of ${\cal H}$, and we define ${\cal H}^{\it II}$ as
the orthogonal complement of ${\cal H}^{\it I}$ with respect to the
scalar product $V_0[f, g]$. Any set of $n$ observables satisfying
\eqref{centred} can then be written as
\begin{equation}
  \label{decompose}
  \obs_i = \obs^{\it I}_i + \obs^{\it II}_i
\end{equation}
with $\obs^{\it I}_i \in {\cal H}^{\it I}$, $\obs^{\it II}_i \in {\cal
  H}^{\it II}$. Further decomposing $\obs^{\it I}_i = \sum_j a_{ij}
{\cal D}_{j}$ and using the constraint \eqref{normalised} we obtain
$a_{ij} = \delta_{ij}$, i.e.\
\begin{equation}
  \label{first}
  \obs^{\it I}_i = {\cal D}_i  \eqpt
\end{equation}
Finally, we have from \eqref{simpleCov}, \eqref{decompose},
\eqref{first}
\begin{equation}
  \label{decomposeCov}
  V(g)_{ij} = N^{-1} V_0[{\cal D}_i, {\cal D}_j] +
              N^{-1} V_0[\obs^{\it II}_i, \obs^{\it II}_j]  \eqpt
\end{equation}
The first term gives the error on the couplings for the optimal
observables ${\cal D}_i$, which is minimal. If the observables
$\obs_i$ have minimal error, too, the second term must be zero, so
that for each $i$ we have $V_0[\obs^{\it II}_i, \obs^{\it II}_i] = 0$
and thus
\begin{equation}
  \label{second}
  \obs^{\it II}_i = 0  \eqcm
\end{equation}
which completes our proof.

In sec.~\ref{sec:simultan} we mentioned that instead of $\mean_i$ one
may use the product $N \mean_i$ measured with fixed luminosity to
extract the couplings \cite{OldObs}. By an argument analogous to the
one of this appendix one finds that up to linear reparametrisations
our observables \eqref{newOptimal} are again the only optimal ones. In
this case linear reparametrisations have to be homogeneous, i.e.\ one
must have $b_i = 0$ in \eqref{reparam}, since adding constants to the
observables can change the induced errors on the coupling parameters.

\end{document}